\documentclass[useAMS,usegraphicx,usenatbib]{mn2e}          

\usepackage{times}

\newcommand{\Msun}{\ensuremath{\,{\rm M}_\odot}}                  
\newcommand{\Rsun}{\ensuremath{\,{\rm R}_\odot}}                  
\newcommand{\Teff}{\ensuremath{T_{\rm eff}}}                      
\newcommand{\logg}{\ensuremath{\log g}}                           
\newcommand{\Mjup}{\ensuremath{\,{\rm M}_{\rm Jup}}}              
\newcommand{\Rjup}{\ensuremath{\,{\rm R}_{\rm Jup}}}              
\newcommand{\Teq}{\ensuremath{T_{\rm eq}^{\,\prime}}}             
\newcommand{\safronov}{\ensuremath{\Theta}}                       
\newcommand{\ms}{\,m\,s$^{-1}$}                                   
\newcommand{\mss}{\,m\,s$^{-2}$}                                  
\newcommand{\as}{\ensuremath{^{\prime\prime}}}                    
\newcommand{\am}{\ensuremath{^\prime}}                            
\newcommand{\FeH}{\ensuremath{\left[\frac{\rm Fe}{\rm H}\right]}} 
\newcommand{\pjup}{\ensuremath{\,\rho_{\rm Jup}}}                 
\newcommand{\psun}{\ensuremath{\,\rho_\odot}}                     
\newcommand{\chir}{\ensuremath{\chi_\nu^{\,2}}}                   
\newcommand{\mc}[1]{\multicolumn{2}{c}{#1}}
\newcommand{\mcc}[1]{\multicolumn{3}{c}{#1}}
\newcommand{\er}[3]{\ensuremath{#1^{+#2}_{-#3}}}
\newcommand{\erc}[3]{\mc{\ensuremath{#1^{+#2}_{-#3}}}}

\newcommand{\ermcc}[5]{\mcc{\ensuremath{{#1\,^{+#2}_{-#3}}\,^{+#4}_{-#5}}}}
\newcommand{\reff}[1]{{#1}}

\title[Defocussed photometry of WASP-15 and WASP-16]
      {High-precision photometry by telescope defocussing. V. WASP-15 and WASP-16%
      \thanks{Based on data collected with the Gamma Ray Burst Optical and Near-Infrared Detector (GROND) at the MPG/ESO 2.2\,m telescope and by MiNDSTEp with the Danish 1.54\,m telescope at the ESO La Silla Observatory.}}

\author[Southworth et al.]
       {John Southworth\,$^{1}$, L.\ Mancini\,$^{2,3}$, P.\ Browne\,$^{4}$, M.\ Burgdorf\,$^{5}$, S.\ Calchi Novati\,$^{6,3}$, 
        \newauthor
        M.\ Dominik\,$^{4}$\thanks{Royal Society University Research Fellow}, 
        T.\ Gerner\,$^{7}$, T.\ C.\ Hinse\,$^{8}$, U.\ G.\ J{\o}rgensen\,$^{9,10}$, N.\ Kains\,$^{11}$, D.\ Ricci\,$^{12}$, 
        \newauthor
        S.\ Sch\"afer\,$^{13}$, F.\ Sch\"onebeck\,$^{7}$, J.\ Tregloan-Reed\,$^{1}$, K.\ A.\ Alsubai\,$^{14}$, V.\ Bozza\,$^{3,15}$,
        \newauthor
        G.\ Chen\,$^{2,16}$, P.\ Dodds\,$^{4}$, S.\ Dreizler\,$^{13}$, X.-S.\ Fang\,$^{17}$, F.\ Finet\,$^{18}$, S.-H.\ Gu\,$^{17}$, S.\ Hardis\,$^{9,10}$,
        \newauthor
        K.\ Harps{\o}e\,$^{9,10}$, Th.\ Henning\,$^{2}$, M.\ Hundertmark\,$^{4}$, J.\ Jessen-Hansen\,$^{19}$, E.\ Kerins\,$^{20}$,
        \newauthor
        H.\ Kjeldsen\,$^{19}$, C.\ Liebig\,$^{4}$, M.\ N.\ Lund\,$^{19}$, M.\ Lundkvist\,$^{19}$, M.\ Mathiasen\,$^{9,10}$,
        \newauthor
        N.\ Nikolov\,$^{2,21}$, M.\ T.\ Penny\,$^{22}$, S.\ Proft\,$^{7}$, S.\ Rahvar\,$^{23}$, K.\ Sahu\,$^{24}$, G.\ Scarpetta\,$^{6,3,15}$,
        \newauthor
        J.\ Skottfelt\,$^{9,10}$, C.\ Snodgrass\,$^{25}$, J.\ Surdej\,$^{17}$, O.\ Wertz\,$^{17}$
        \\
        $^{1}$\,Astrophysics Group, Keele University, Staffordshire, ST5 5BG, UK \\
        $^{2}$\,Max Planck Institute for Astronomy, K\"onigstuhl 17, 69117 Heidelberg, Germany \\
        $^{3}$\,Dipartimento di Fisica ``E. R. Caianiello'', Universit\`a di Salerno, Via Ponte Don Melillo, 84084-Fisciano (SA), Italy \\
        $^{4}$\,SUPA, University of St Andrews, School of Physics \& Astronomy, North Haugh, St Andrews, KY16 9SS, UK \\
        $^{5}$\,HE Space Operations GmbH, Flughafenallee 24, D-28199 Bremen, Germany \\
        $^{6}$\,Istituto Internazionale per gli Alti Studi Scientifici (IIASS), 84019 Vietri Sul Mare (SA), Italy \\
        $^{7}$\,Zentrum f\"ur Astronomie, Universit\"at Heidelberg, M\"onchhofstra{\ss}e 12-14, 69120 Heidelberg, Germany \\
        $^{8}$\,Korea Astronomy and Space Science Institute, Daejeon 305-348, Republic of Korea \\
        $^{9}$\,Niels Bohr Institute, K{\o}benhavns Universitet, Juliane Maries vej 30, 2100 Copenhagen \O, Denmark \\
        $^{10}$\,Centre for Star and Planet Formation, Natural History Museum of Denmark, K{\o}benhavns Universitet, {\O}ster Voldgade 5-7, 1350 K{\o}benhavn K, Denmark \\
        $^{11}$\,European Southern Observatory, Karl-Schwarzschild-Stra{\ss}e 2, 85748 Garching bei M\"unchen, Germany \\
        $^{12}$\,Instituto de Astronom\'ia -- UNAM, Km 103 Carretera Tijuana Ensenada, 422860, Ensenada (Baja Cfa), Mexico \\
        $^{13}$\,Institut f\"ur Astrophysik, Georg-August-Universit\"at G\"ottingen, Friedrich-Hund-Platz 1, 37077 G\"ottingen, Germany \\
        $^{14}$\,Qatar Foundation, PO Box 5825, Doha, Qatar \\
        $^{15}$\,Istituto Nazionale di Fisica Nucleare, Sezione di Napoli, Napoli, Italy \\
        $^{16}$\,Purple Mountain Observatory \& Key Laboratory for Radio Astronomy, Chinese Academy of Sciences, 2 West Beijing Road, Nanjing 210008, China \\
        $^{17}$\,National Astronomical Observatories/Yunnan Observatory, Chinese Academy of Sciences, Kunming 650011, China \\
        $^{18}$\,Institut d'Astrophysique et de G\'eophysique, Universit\'e de Li\`ege, 4000 Li\`ege, Belgium \\
        $^{19}$\,Stellar Astrophysics Centre (SAC), Department of Physics and Astronomy, Aarhus University, Ny Munkegade 120, DK-8000 Aarhus C, Denmark \\
        $^{20}$\,Jodrell Bank Centre for Astrophysics, University of Manchester, Oxford Road, Manchester M13 9PL, UK \\
        $^{21}$\,Astrophysics Group, University of Exeter, Stocker Road, EX4 4QL, Exeter, UK \\
        $^{22}$\,Department of Astronomy, Ohio State University, 140 W. 18th Ave., Columbus, OH 43210, USA \\
        $^{23}$\,Department of Physics, Sharif University of Technology, P.\,O.\,Box 11155-9161 Tehran, Iran \\
        $^{24}$\,Space Telescope Science Institute, 3700 San Martin Drive, Baltimore, MD 21218, USA \\
        $^{25}$\,Max-Planck-Institute for Solar System Research, Max-Planck Str.\ 2, 37191 Katlenburg-Lindau, Germany 
}

\begin{document} \maketitle 

\clearpage

\begin{abstract}
We present new photometric observations of WASP-15 and WASP-16, two transiting extrasolar planetary systems \reff{with measured orbital obliquities but without} photometric follow-up since their discovery papers. Our new data for WASP-15 comprise observations of one transit simultaneously in four optical passbands using GROND on the MPG/ESO 2.2\,m telescope, plus coverage of half a transit from DFOSC on the Danish 1.54\,m telescope, both at ESO La Silla. For WASP-16 we present observations of four complete transits, all from the Danish telescope. We use these new data to refine the measured physical properties and orbital ephemerides of the two systems. Whilst our results are close to the originally-determined values for WASP-15, we find that the star and planet in the WASP-16 system are both larger and less massive than previously thought. 
\end{abstract}



\section{Introduction}                                                                                                              \label{sec:intro}

The number of known transiting extrasolar planets (TEPs) is rapidly increasing, and currently stands at 310\footnote{Data taken from the Transiting Extrasolar Planet Catalogue (TEPCat) available at: {\tt http://www.astro.keele.ac.uk/jkt/tepcat/}}. Their diversity is also escalating: the radius of the largest known example is 40 times greater than that of the smallest. There is a variation of over three orders of magnitude in their masses, excluding those without mass measurements and those which are arguably brown dwarfs. Whilst a small subset of this population has been extensively investigated, the characterisation of the majority is limited to modest photometry and spectroscopy presented in their discovery papers. 

The bottleneck in our understanding of the physical properties of most TEPs is the quality of the available transit light curves, which are of fundamental importance for measuring the stellar density and the ratio of the radius of the planet to that of the star. Additional contributions, which arise from the spectroscopic parameters of the host star and the constraints on its physical properties from theoretical models, are usually dwarfed by the uncertainties in the photometric parameters derived from the light curves.

We are therefore undertaking a project aimed at characterising TEPs visible from the Southern hemisphere (see \citealt{Me+12mn3} and references therein), by obtaining high-precision light curves of their transits. We use the telescope defocussing technique, discussed in detail in \citet{Me+09mn}, to collect photometric measurements with very low levels of Poisson and correlated noise. This method is able to achieve light curves of remarkable precision \citep[e.g.][]{TregloanSouthworth13mn}. In this work we present new observations and determinations of the physical properties of WASP-15 and WASP-16, based on nine light curves covering six transits in total.

                                                                                                                            \label{sec:intro:history}
\subsection{Case history}

WASP-15 was identified as a TEP by \citet{West+09aj}, who found it to be a low-density object ($\rho_2 = 0.186 \pm 0.026$\pjup) orbiting a slightly evolved and comparatively hot host star ($\Teff = 6300 \pm 100$\,K). Other measurements of the effective temperature of the host star have been made by \citet*{Maxted++11mn}, who found $\Teff = 6210 \pm 60$\,K from the infrared flux method \citep{Blackwell++80aa}, and by \citet{Doyle+13mn}, whose detailed spectroscopic analysis yielded $\Teff = 6405 \pm 80$\,K. 

\citet{Triaud+10aa} observed the Rossiter-McLaughlin effect for WASP-15 and found the system to exhibit significant obliquity: the sky-projected angle between the rotational axis of the host star and the orbital axis of the planet is $\lambda = \er{139.6}{4.3}{5.2}$\,degrees. This is consistent with previous findings that misaligned planets are found only around hotter stars \citep{Winn+10apj3}, although tidal effects act to align them over time \citep{Triaud11aa,Albrecht+12apj2}.

The discovery of the planetary nature of WASP-16 was made by \citet{Lister+09apj}, who characterised it as a Jupiter-like planet orbiting a star similar to our Sun. \citet{Maxted++11mn} and \citet{Doyle+13mn} measured the host star's \Teff\ to be $5550 \pm 60$\,K and $5630 \pm 70$\,K, respectively, in mutual agreement and a little cooler than the value of $5700 \pm 150$\,K found in the discovery paper.

Observations of the Rossiter-McLaughlin effect for WASP-16 have yielded obliquities consistent with zero: \citet{Brown+12mn} measured $\lambda = \er{11}{26}{19}$\,degrees and \citet{Albrecht+12apj2} found $\lambda = \er{-4}{11}{14}$\,degrees. The large uncertainties in these assessments are due to the low rotational velocity of the star, which results in a small amplitude for the Rossiter-McLaughlin effect.

The physical properties of both systems were comparatively ill-defined, as they rested on few dedicated follow-up light curves: only one light curve in the case of WASP-16, and two datasets afflicted with correlated noise in the case of WASP-15. All three datasets were obtained using EulerCam on the 1.2\,m Swiss Euler telescope at ESO La Silla. In this work we present the first follow-up photometry since the discovery paper for both systems, totalling nine new light curves covering six transits. This new material has allowed us to significantly improve the precision of the measured physical properties. Our analysis also benefited from refined constraints on the atmospheric characteristics of the host stars, as discussed above.


\section{Observations and data reduction}                                                                                             \label{sec:obs}

\begin{table*} \centering
\caption{\label{tab:obslog} Log of the observations presented in this work. $N_{\rm obs}$ is the number
of observations, $T_{\rm exp}$ is the exposure time, $T_{\rm obs}$ is the observational cadence, 
and  `Moon illum.' is the fractional illumination of the Moon at the midpoint of the transit.}
\begin{tabular}{lccccccccccc} \hline
Transit & Date of   & Start time & End time  &$N_{\rm obs}$ & $T_{\rm exp}$ & $T_{\rm obs}$ & Filter & Airmass & Moon & Aperture   & Scatter \\
        & first obs &    (UT)    &   (UT)    &              & (s)           & (s)           &        &         &illum.& radii (px) & (mmag)  \\
\hline
\multicolumn{10}{l}{WASP-15:} \\
DFOSC & 2010 06 09 & 23:09 & 03:09 &  92 & 120    & 155 & Bessell $R$ & 1.15 $\to$ 1.00 $\to$ 1.08 & 0.068 & 32, 45, 70  & 0.492 \\ 
GROND & 2012 04 19 & 02:23 & 09:39 & 229 & 62--45 & 115 & Gunn $g$    & 1.17 $\to$ 1.00 $\to$ 2.09 & 0.040 & 50, 75, 95  & 0.640 \\ 
GROND & 2012 04 19 & 02:23 & 09:39 & 228 & 62--45 & 115 & Gunn $r$    & 1.17 $\to$ 1.00 $\to$ 2.09 & 0.040 & 50, 75, 95  & 0.481 \\ 
GROND & 2012 04 19 & 02:23 & 09:39 & 225 & 62--45 & 115 & Gunn $i$    & 1.17 $\to$ 1.00 $\to$ 2.09 & 0.040 & 50, 75, 100 & 0.607 \\ 
GROND & 2012 04 19 & 02:23 & 09:39 & 227 & 62--45 & 115 & Gunn $z$    & 1.17 $\to$ 1.00 $\to$ 2.09 & 0.040 & 50, 75, 100 & 0.725 \\ 
\multicolumn{10}{l}{WASP-16:} \\
DFOSC & 2010 05 10 & 01:33 & 06:17 & 131 & 100    & 128 & Bessell $R$ & 1.18 $\to$ 1.01 $\to$ 1.22 & 0.156 & 30, 50, 80  & 0.542 \\ 
DFOSC & 2010 06 28 & 23:25 & 04:10 & 136 & 75     & 102 & Bessell $R$ & 1.05 $\to$ 1.01 $\to$ 1.55 & 0.937 & 30, 40, 60  & 1.294 \\ 
DFOSC & 2011 05 13 & 01:07 & 05:42 & 140 & 90     & 118 & Bessell $R$ & 1.23 $\to$ 1.01 $\to$ 1.15 & 0.752 & 26, 40, 60  & 0.586 \\ 
DFOSC & 2011 07 01 & 23:18 & 04:36 & 160 & 90     & 120 & Bessell $R$ & 1.05 $\to$ 1.01 $\to$ 1.88 & 0.006 & 34, 45, 70  & 0.670 \\ 
\hline \end{tabular} \end{table*}

\begin{figure} \includegraphics[width=0.48\textwidth,angle=0]{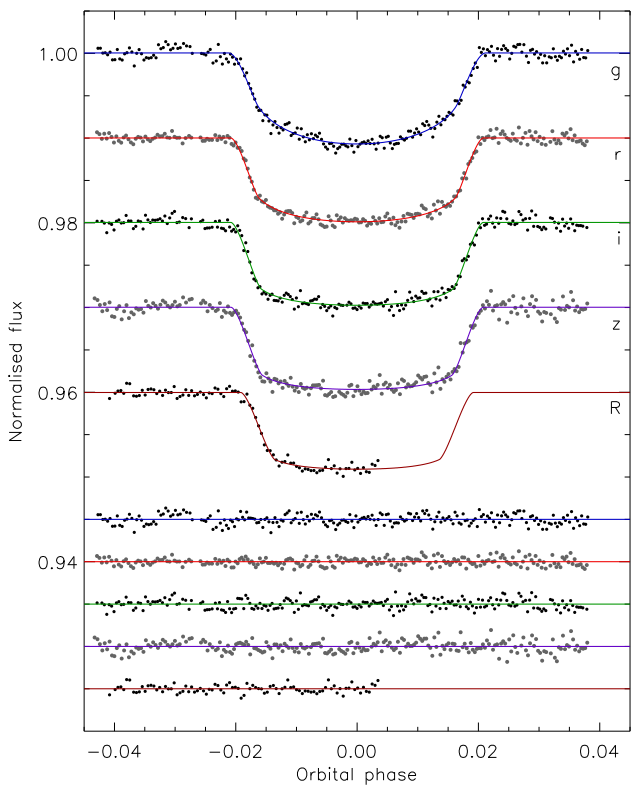}
\caption{\label{fig:griz} Optical light curves of WASP-15. The first four are from 
GROND and the fifth is from DFOSC. The {\sc jtkebop} best fit is shown for each datasets, 
and the residuals of the fit are plotted near the base of the figure.} \end{figure}

\begin{figure} \includegraphics[width=0.48\textwidth,angle=0]{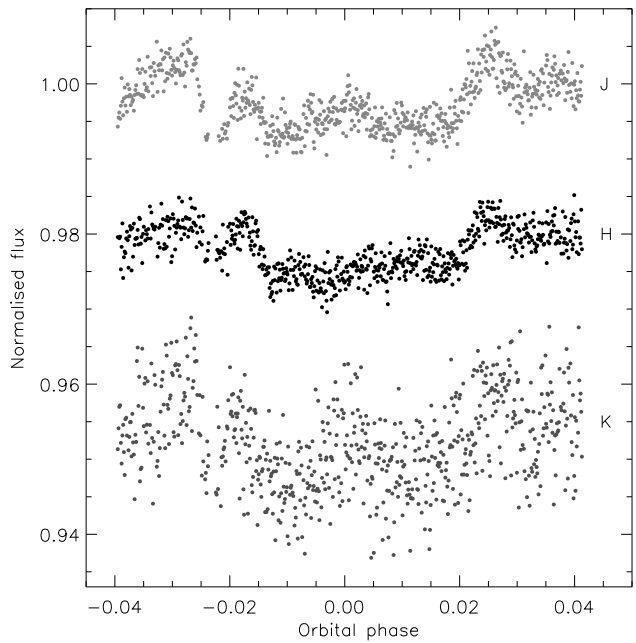}
\caption{\label{fig:jhk} Near-IR light curves of WASP-15 from GROND. The passbands
are labelled on the right of the figure.} \end{figure}

\begin{figure} \includegraphics[width=0.48\textwidth,angle=0]{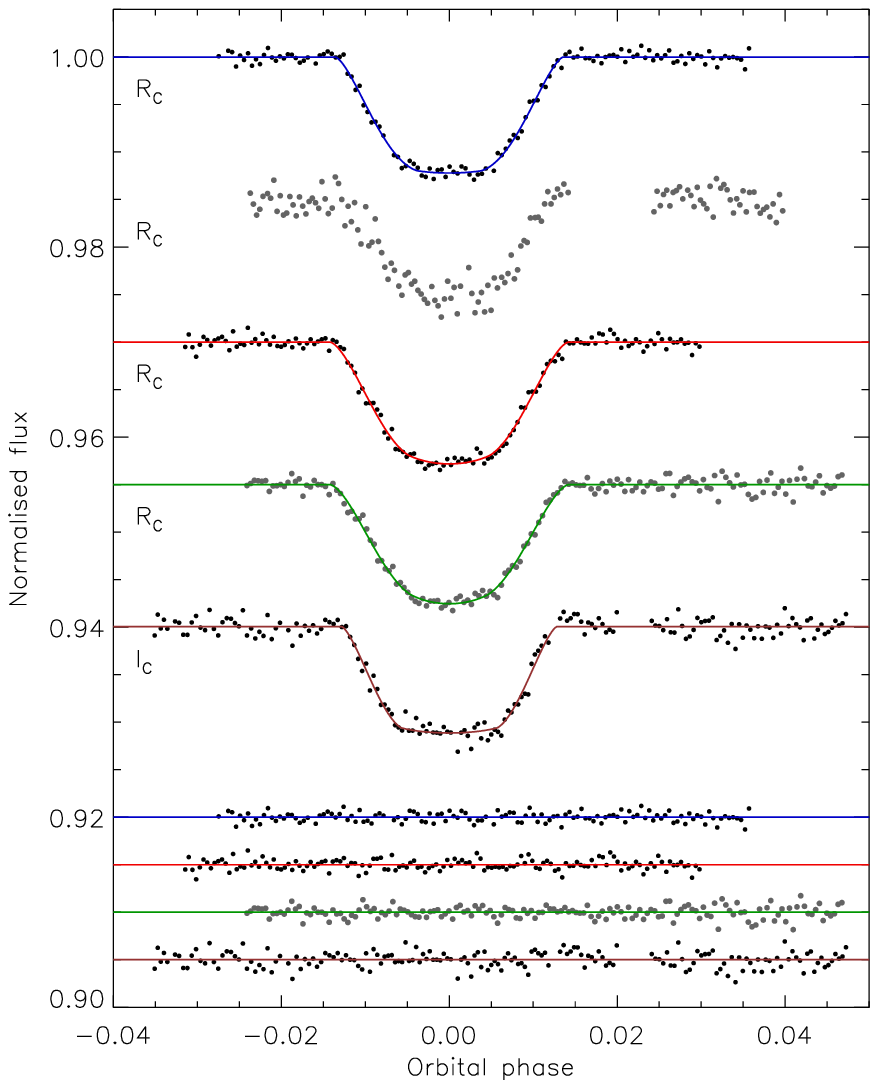}
\caption{\label{fig:w16} The four new light curves of WASP-16, plotted in the 
order they are given in Table\,\ref{tab:obslog}, plus a fifth dataset (lower 
curve) from  \citet{Lister+09apj}. The second dataset is unreliable and a 
fitted model is not plotted for it. The {\sc jktebop} best fits for the 
other datasets are shown as solid lines and the residuals of the fits  
are plotted near the base of the figure.} \end{figure}

We observed one transit of WASP-15 on the night of 2012/01/19 using the GROND instrument mounted on the MPG/ESO 2.2\,m telescope at La Silla, Chile. The field of view of this instrument is 5.4$^{\prime}$$\times$5.4$^{\prime}$ at a plate scale of 0.158$^{\prime\prime}$\,px$^{-1}$. Observations were obtained simultaneously in the $g$, $r$, $i$ and $z$ passbands and covered a full transit plus significant time intervals before ingress and after egress. CCD readout occurred in slow mode. The telescope was defocussed and we autoguided throughout the observations. The moon was below the horizon during the observing sequence. An observing log is given in Table\,\ref{tab:obslog}.

The data were reduced with the {\sc idl}\footnote{The acronym {\sc idl} stands for Interactive Data Language and is a trademark of ITT Visual Information Solutions. For further details see: {\tt http://www.ittvis.com/ProductServices/IDL.aspx}.} pipeline described by \citet{Me+09mn}, which uses the {\sc daophot} package \citep{Stetson87pasp} to perform aperture photometry with the {\sc aper}\footnote{{\sc aper} is part of the {\sc astrolib} subroutine library distributed by NASA. For further details see: {\tt http://idlastro.gsfc.nasa.gov/}.} routine. The apertures were placed by hand and the stars were tracked by cross-correlating each image against a reference image. We tried a wide range of aperture sizes and retained those which gave photometry with the lowest scatter compared to a fitted model. In line with previous experience, we found that the shape of the light curve is very insensitive to the aperture sizes.

We calculated differential-photometry light curves of our target star by combining all good comparison stars into an ensemble with weights optimised to minimise the scatter of the observations taken outside transit. We rectified the data to a zero-magnitude baseline by subtracting a second-order polynomial whose coefficients were optimised simultaneously with the weights of the comparison stars. The effect of this normalisation was subsequently taken into account when modelling the data. The final GROND optical light curves are shown in Fig.\,\ref{fig:griz}. Our timestamps were converted to the BJD(TDB) timescale \citep{Eastman++10pasp}.

We also used GROND to obtain photometry in the $J$, $H$ and $K$ passbands simultaneously with the optical observations. The field of view of the GROND near-infrared channels is 10$^{\prime}$$\times$10$^{\prime}$ at a plate scale of 0.60$^{\prime\prime}$\,px$^{-1}$. These were reduced following standard techniques and with trying multiple alternative approaches to decorrelate the data against airmass and centroid position of the target star. We were unable to obtain good light curves from these data, and suspect that this is because the brightness of WASP-15 pushed the pixel count rates into the nonlinear regime, causing the systematic noise which is obvious in Fig.\,\ref{fig:jhk}. 

A transit of WASP-15 was also observed using the DFOSC imager on board the 1.54\,m Danish Telescope at La Silla, which has a field of view of 13.7\am$\times$13.7\am\ and a plate scale of 0.39\as\,pixel$^{-1}$. We defocussed the telescope and autoguided. Several images were taken prior to the main body of observations in order to check for faint nearby stars which might contaminate the point spread function (PSF) of our target star, and none were found. Unfortunately high winds forced the closure of the dome shortly after the midpoint of the transit, which has limited the usefulness of these data. The data were reduced as above, except that a first-order polynomial (i.e.\ a straight line) was used as the function to rectify the light curve to zero differential magnitude.

Four transits of WASP-16 were obtained using the DFOSC imager and the same approach as for the WASP-15 transit above. Three of the transits were observed in excellent weather conditions whilst the moon was below the horizon, and these yield excellent light curves. The data were reduced as above, using a straight-line fit to the out-of-transit data. A small number of images taken in focus showed that there are faint stars separated by 32 and 45 pixels from the centre of the PSF of WASP-16. They are fainter than our target star by more than 8.7 and 6.8 mag, respectively, so have a negligible effect on our results.

The second transit was undermined by non-photometric conditions, bright moonlight, and a computer crash shortly after the transit finished. This transit is shallower than the other three, and we attribute this to a count rate during the observing sequence that became sufficiently high to enter the regime of significant nonlinearity in the CCD response. The data for this transit were not included in subsequent analyses. All four light curves are shown in Fig.\,\ref{fig:w16}, along with the Euler Telescope data from \citet{Lister+09apj}. All our reduced data will be made available at the CDS\footnote{{\tt http://vizier.u-strasbg.fr/}}.


\section{Light curve analysis}                                                                                                         \label{sec:lc}

The analysis of our light curves was performed using the {\it Homogeneous Studies} methodology (see \citealt{Me12mn} and references therein). The light curves were modelled using the {\sc jktebop}\footnote{{\sc jktebop} is written in {\sc fortran77} and the source code is available at {\tt http://www.astro.keele.ac.uk/jkt/codes/jktebop.html}} code \citep{Me++04mn}, which represents the star and planet as biaxial spheroids. The main parameters of the model are the fractional radii of the star and planet, $r_{\rm A}$ and $r_{\rm b}$, and the orbital inclination, $i$. The fractional radii are the true radii of the objects divided by the orbital semimajor axis. They were parameterised by their sum and ratio:
$$ r_{\rm A} + r_{\rm b} \qquad \qquad k = \frac{r_{\rm b}}{r_{\rm A}} = \frac{R_{\rm b}}{R_{\rm A}} $$
as the latter are less strongly correlated than the fractional radii themselves.

\subsection{Orbital period determination}                                                                                        \label{sec:lc:porb}

\begin{table} \begin{center}
\caption{\label{tab:minima} Times of minimum light of WASP-15 (upper) and 
WASP-16 (lower) and their residuals versus the ephemeris derived in this work.
\newline {\bf References:}
(1) \citet{West+09aj};
(2) T.\,G.Tan (ETD);
(3) This work;
(4) \citet{Lister+09apj};
(5) E.\ Fernandez-Lajus, Y.\ Miguel, A.\ Fortier \& R.\ Di Sisto (TRESCA);
(6) M.\ Vra\v{s}\'t\'ak (TRESCA);
(7) M.\ Schneiter, C.\ Colazo \& P.\ Guzzo (TRESCA);
(8) F.\ Tifner (TRESCA).}
\begin{tabular}{l@{\,$\pm$\,}l r r l} \hline
\multicolumn{2}{l}{Time of minimum}   & Cycle  & Residual & Reference \\
\multicolumn{2}{l}{(BJD(TDB) $-$ 2400000)} & number & (JD)    &           \\
\hline
54584.69860 & 0.00029 &     0.0 &  0.00001 & 1        \\      
55320.10914 & 0.00135 &   196.0 & -0.00055 & 2        \\      
56036.75990 & 0.00028 &   387.0 & -0.00042 & 3 ($g$)  \\      
56036.76049 & 0.00019 &   387.0 &  0.00017 & 3 ($r$)  \\      
56036.76044 & 0.00023 &   387.0 &  0.00012 & 3 ($i$)  \\      
56036.76020 & 0.00028 &   387.0 & -0.00012 & 3 ($z$)  \\      
\hline
54584.42915 & 0.00029 &     0.0 &  0.00017 & 4        \\      
55276.75911 & 0.00036 &   222.0 & -0.00057 & 5        \\      
55311.06859 & 0.00185 &   233.0 &  0.00424 & 2        \\      
55314.18358 & 0.00100 &   234.0 &  0.00062 & 2        \\      
55326.65793 & 0.00019 &   238.0 &  0.00054 & 3        \\      
55376.55453 & 0.00049 &   254.0 & -0.00056 & 3        \\      
55688.41629 & 0.00177 &   354.0 &  0.00052 & 6        \\      
55694.65194 & 0.00020 &   356.0 & -0.00104 & 3        \\      
55744.55044 & 0.00023 &   372.0 & -0.00025 & 3        \\      
56037.70089 & 0.00024 &   466.0 &  0.00117 & 7        \\      
56087.59554 & 0.00102 &   482.0 & -0.00189 & 8        \\      
\hline \end{tabular} \end{center} \end{table}   

\begin{figure*} \includegraphics[width=\textwidth,angle=0]{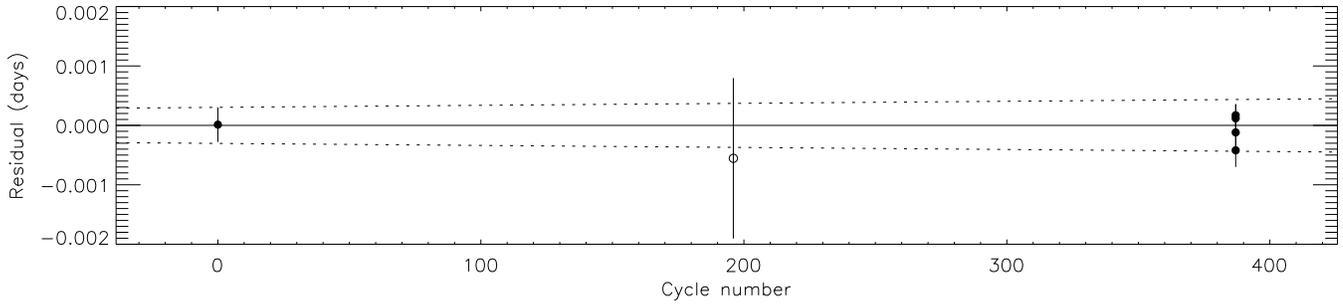}
\caption{\label{fig:minima15} Plot of the residuals of the timings of
mid-transit of WASP-15 versus a linear ephemeris. Timings obtained from 
amateur observations are plotted using open circles, and other timings 
are plotted with filled circles. \reff{The dotted lines show the total 
1$\sigma$ uncertainty in the ephemeris as a function of cycle number.}} 
\end{figure*}

\begin{figure*} \includegraphics[width=\textwidth,angle=0]{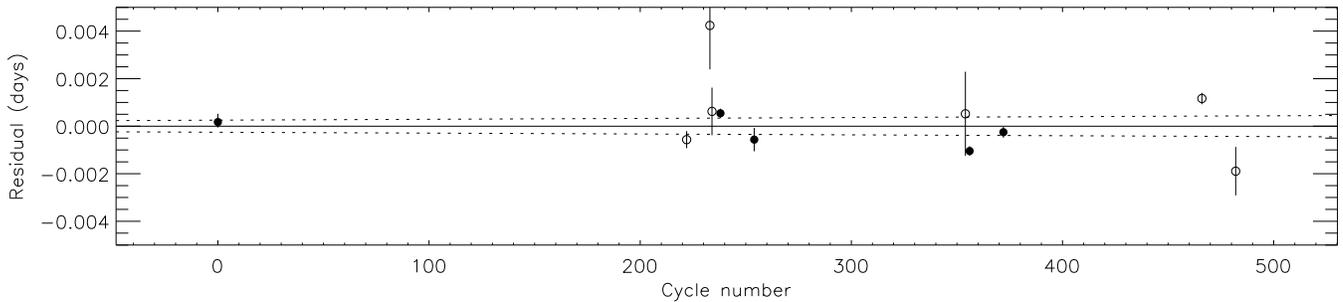}
\caption{\label{fig:minima16} Plot of the residuals of the timings of
mid-transit of WASP-16 versus a linear ephemeris. Other comments are
the same as for Fig\,\ref{fig:minima15}.} \end{figure*}

Our first step was to obtain refined orbital ephemerides. Each of our transit light curves was fitted individually and their errorbars rescaled to give $\chi^2_\nu = 1.0$ versus the fitted model. This is needed as the uncertainties from the {\sc aper} photometry algorithm tend to be underestimated. We then fitted the revised datasets and ran Monte Carlo simulations to measure the transit midpoints with robust errorbars. 

Our own times of transit midpoint were supplemented with those from the discovery papers \citep{West+09aj,Lister+09apj}. The reference times of transit ($T_0$) from these papers are given on the BJD and HJD time conventions, respectively, but the timescales these refer to are not specified \citep[see][]{Eastman++10pasp}. D.\ R.\ Anderson (private communication) has confirmed that the timescales used in these, and the other early WASP planet discovery papers, is UTC. We therefore converted the timings to TDB.

We also compiled publicly available measurements from the Exoplanet Transit Database (ETD\footnote{The Exoplanet Transit Database (ETD) can be found at: {\tt http://var2.astro.cz/ETD/credit.php}}), which makes available datasets from amateur observers affiliated with TRESCA\footnote{The TRansiting ExoplanetS and CAndidates (TRESCA) website can be found at: {\tt http://var2.astro.cz/EN/tresca/index.php}}. We retained only those timing measurements based on light curves where all four contact points of the transit are easily identifiable. We assumed that the times were all on the UTC timescales and converted them to TDB for congruency with our own data.

Once the available times of mid-transit had been assembled, we fitted them with straight lines to determine new orbital ephemerides. Table\,\ref{tab:minima} reports all times of mid-transit used for both objects, plus the residuals versus a linear ephemeris. The new ephemeris for WASP-15 is:
$$ T_0 = {\rm BJD(TDB)} \,\, 2\,454\,584.69859 (29) \, + \, 3.75209748 (81) \times E $$
where $E$ represents the cycle count with respect to the reference epoch and the bracketed quantities show the uncertainty in the final digit of the preceding number. The reduced $\chi^2$ of the fit to the timings is encouragingly small at $\chir = 0.78$ \reff{for four degrees of freedom}, which suggests the orbital period is constant and the uncertainties of the available times of minimum are reasonable. A plot of the fit is shown in Fig.\,\ref{fig:minima15}.

The situation for WASP-16 is less favourable, with $\chir = 2.59$ \reff{(nine degrees of freedom)}, and large residuals for several of the most precise datapoints (Fig.\,\ref{fig:minima16}). We have reason to be cautious about our own timings, as the DFOSC timestamps are known to have been incorrect for the 2009 season \citep{Me+09apj}. This issue was minimised for the 2010 season (which contains the first two transits of WASP-16 we observed) and fixed for the 2011 season (which contains the third and fourth WASP-16 transits), so the disagreement between the two 2011 transits cannot currently be dismissed as an instrumental effect. WASP-16 should be monitored in the future to investigate the possibility that it undergoes transit timing variations. In the meantime, the linear ephemeris given by the timings in Table\,\ref{tab:minima} is:
$$ T_0 = {\rm BJD(TDB)} \,\, 2\,454\,584.42898 (38) \, + \, 3.1186068 (12) \times E $$
where the errorbars have been multiplied by $\sqrt{2.59}$ to account for the large \chir.

\subsection{Light curve modelling}                                                                                                  \label{sec:lc:lc}

\begin{table*} \caption{\label{tab:15:lcfit} Parameters of the fit to the light curves of 
WASP-15 from the {\sc jktebop} analysis (top). The final parameters are given in bold and the 
parameters found by other studies are shown (below). Quantities without quoted uncertainties 
were not given by those authors but have been calculated from other parameters which were.}
\begin{tabular}{l r@{\,$\pm$\,}l r@{\,$\pm$\,}l r@{\,$\pm$\,}l r@{\,$\pm$\,}l r@{\,$\pm$\,}l}
\hline 
Source       & \mc{$r_{\rm A}+r_{\rm b}$} & \mc{$k$} & \mc{$i$ ($^\circ$)} & \mc{$r_{\rm A}$} & \mc{$r_{\rm b}$} \\
\hline
GROND $g$ & 0.1538 & 0.0121 & 0.0967 & 0.0031 & 85.56 & 1.21 & 0.1403 & 0.0106 & 0.01356 & 0.00137 \\
GROND $r$ & 0.1466 & 0.0065 & 0.0936 & 0.0014 & 86.13 & 0.72 & 0.1341 & 0.0058 & 0.01255 & 0.00070 \\
GROND $i$ & 0.1505 & 0.0063 & 0.0956 & 0.0013 & 85.68 & 0.63 & 0.1374 & 0.0057 & 0.01313 & 0.00069 \\
GROND $z$ & 0.1519 & 0.0084 & 0.0959 & 0.0014 & 85.51 & 0.84 & 0.1386 & 0.0076 & 0.01329 & 0.00080 \\
DFOSC $R$ & 0.1545 & 0.0086 & 0.0933 & 0.0023 & 85.27 & 0.87 & 0.1413 & 0.0078 & 0.01318 & 0.00093 \\
\hline
Final results&{\bf0.1500}&{\bf0.0037}&{\bf0.09508}&{\bf0.00078}&{\bf85.74}&{\bf0.38}&{\bf0.1370}&{\bf0.0033}&{\bf0.01303}&{\bf0.00039}\\
\hline
\citet{West+09aj} & \mc{0.1436} & 0.099 & 0.001 & \mc{ } & \mc{0.1331} & \mc{0.01318} \\
\citet{Triaud+10aa} & \mc{0.1474} & \erc{0.09842}{0.00067}{0.00058} & \erc{85.96}{0.29}{0.41} & \erc{0.1342}{0.0039}{00027} & \erc{0.01321}{0.00047}{0.00030} \\
\hline \end{tabular} \end{table*}

\begin{table*} \caption{\label{tab:16:lcfit} Parameters of the fit to the light curves of WASP-16 
from the {\sc jktebop} analysis (top). The final parameters are given in bold and the parameters 
found by \citet{Lister+09apj} are shown below. Quantities without quoted uncertainties were not 
given by \citet{Lister+09apj} but have been calculated from other parameters which were.}
\begin{tabular}{l r@{\,$\pm$\,}l r@{\,$\pm$\,}l r@{\,$\pm$\,}l r@{\,$\pm$\,}l r@{\,$\pm$\,}l}
\hline 
Source       & \mc{$r_{\rm A}+r_{\rm b}$} & \mc{$k$} & \mc{$i$ ($^\circ$)} & \mc{$r_{\rm A}$} & \mc{$r_{\rm b}$} \\
\hline
DFOSC transit 1 & 0.1365 & 0.0052 & 0.1118 & 0.0060 & 83.84 & 0.39 & 0.1228 & 0.0053 & 0.01364 & 0.00043 \\
DFOSC transit 3 & 0.1354 & 0.0073 & 0.1198 & 0.0032 & 84.14 & 0.59 & 0.1209 & 0.0063 & 0.01448 & 0.00090 \\
DFOSC transit 4 & 0.1362 & 0.0050 & 0.1204 & 0.0035 & 84.09 & 0.44 & 0.1216 & 0.0046 & 0.01464 & 0.00063 \\
Euler transit   & 0.1219 & 0.0071 & 0.1074 & 0.0038 & 84.75 & 0.55 & 0.1101 & 0.0067 & 0.01182 & 0.00059 \\
\hline
Final results&{\bf0.1362}&{\bf0.0031}&{\bf0.1190}&{\bf0.0022}&{\bf83.99}&{\bf0.26}&{\bf0.1218}&{\bf0.0030}&{\bf0.01402}&{\bf0.00033}\\
\hline
\citet{Lister+09apj} & \mc{0.1167} & \erc{0.1095}{0.0023}{0.0018} & \erc{85.22}{0.27}{0.43} & \mc{0.1065} & \mc{0.01012} \\
\hline \end{tabular} \end{table*}
 
We modelled each of our light curves of WASP-15 and WASP-16 individually, using {\sc jktebop} to fit for $r_{\rm A}+r_{\rm b}$, $k$, $i$ and $T_0$. The best-fitting models are shown in Figs.\ \ref{fig:griz} and \ref{fig:w16}. This individual approach was necessary to allow for differing amounts of limb darkening (LD) for WASP-15 and for possible timing variations in WASP-16, and has the advantage of providing an opportunity to assess errorbars by comparing multiple independent sets of results rather than relying on statistical algorithms. The DFOSC transit for WASP-15 lacks coverage of the egress phases so was modelled with $T_0$ fixed at the value predicted by the orbital ephemeris, and the second transit of WASP-16 was ignored due to the systematic errors discussed in Sect.\,\ref{sec:obs}. The follow-up photometry for WASP-15 presented by \citet{West+09aj} was not considered as it contains substantial red noise. The Euler telescope light curve of WASP-16 \citep{Lister+09apj} was added to our analysis as it has full coverage of a transit event with reasonably high precision.

Light curve models were obtained using each of five LD laws \citep[see][]{Me08mn}, with the linear coefficients either fixed at theoretically predicted values\footnote{Theoretical LD coefficients were obtained by bilinear interpolation to the host star's \Teff\ and \logg\ using the {\sc jktld} code available from: {\tt http://www.astro.keele.ac.uk/jkt/codes/jktld.html}} or included as fitted parameters. We made no attempt to fit for both coefficients in the four bi-parametric laws as they are very strongly correlated \citep{Me08mn,Carter+08apj}. The nonlinear coefficients were instead perturbed by $\pm$0.1 on a flat distribution when running the error analysis algorithms, in order to account for their intrinsic uncertainty.

A circular orbit was adopted for both systems as the radial velocities indicate circularity with limits in eccentricity of $e < 0.087$ for WASP-15 \citep{Triaud+10aa} and $e < 0.052$ for WASP-16 \citep{Pont+11mn}. The coefficients of a polynomial function of the out-of-transit magnitude were included when modelling the GROND data, to account for the fact that such a function was used to normalise the data when constructing the differential magnitudes. \reff{We checked for correlations between the coefficients of the polynomial and the other parameters of the fit, finding a significant correlation only between $k$ and the quadratic coefficient. The correlation coefficients in this case are in the region of $0.4$ for the $g$-band), $0.5$ for $r$, and $0.65$ for $i$ and $z$, depending on the specifics of how LD was treated. The uncertainties in the resulting parameters induced by this correlation are accounted for in our methods for estimating the parameter uncertainties.}

Errorbars for the fitted parameters were obtained in two ways: from 1000 Monte Carlo simulations for each solution, and via a residual-permutation algorithm \citep{Me08mn}. The final parameter values are the unweighted mean of those from the solutions involving the four two-parameter LD laws. Their errorbars are the larger of the Monte-Carlo or residual-permutation alternatives, with an extra contribution to account for variations between solutions with the different LD laws. Tables of individual results for each light curve can be found in the Supplementary Information.

For WASP-15, we found that the residual-permutation method returned moderately larger uncertainties for the $g$ and $z$ light curves, as expected from Fig.\,\ref{fig:griz}. We were able to adopt solutions with the linear LD coefficient fitted for the GROND data, but had to use solutions with fixed LD coefficients for the DFOSC observations as they only cover half a transit. The sets of photometric parameters agree extremely well (Table\,\ref{tab:15:lcfit}), and were combined into a weighted mean after downweighting the DFOSC transit by doubling the parameter errorbars. Published results are in acceptable agreement with these weighted mean values.

For WASP-16 the results for the three DFOSC transits agree very well with each other but not with those for the Euler dataset (Table\,\ref{tab:16:lcfit}), which is unsurprising given the best fits plotted in Fig.\,\ref{fig:w16}. We therefore combined only the results from the DFOSC transits into a weighted mean to obtain our final photometric parameters. The errorbars quoted by \citet{Lister+09apj} appear to be rather small given the available data and the discrepancy with our follow-up observations.


\section{Physical properties}

\begin{table} \centering \caption{\label{tab:spec} Spectroscopic properties of 
the host stars in WASP-15 and WASP-16 adopted from the literature and 
used in the determination of the physical properties of the systems.
\newline {\bf References:}
(1) \citet{Doyle+13mn};
(2) \citet{Triaud+10aa};
(3) \citet{Lister+09apj}.}
\begin{tabular}{l r@{\,$\pm$\,}l c r@{\,$\pm$\,}l c}
\hline 
Source      & \mc{WASP-15} & Ref & \mc{WASP-16} & Ref \\
\hline
\Teff\ (K)        & 6405 & 80 & 1   & 5630 & 70 & 1   \\
\FeH\ (dex)       & 0.00 & 0.10 & 1 & 0.07 & 0.10 & 1 \\
$K_{\rm A}$ (\ms) & 64.6 & 1.2 & 2  & 116.7 & 2.2 & 3 \\
\hline \end{tabular} \end{table}
 
\begin{table*} \caption{\label{tab:15:model} Derived physical properties of the WASP-15 system and a comparison to 
previous measurements. Separate statistical and systematic errorbars are given for the results from the current work.}
\begin{tabular}{l r@{\,$\pm$\,}c@{\,$\pm$\,}l r@{\,$\pm$\,}l r@{\,$\pm$\,}l r@{\,$\pm$\,}l r@{\,$\pm$\,}l}
\hline 
\ & \mcc{This work} & \mc{\citet{West+09aj}} & \mc{\citet{Triaud+10aa}} & \mc{\citet{Doyle+13mn}} \\
\hline
$M_{\rm A}$    (\Msun) & 1.305    & 0.051    & 0.006     & 1.18 & 0.12         & \erc{1.18}{0.14}{0.12}       & 1.23 & 0.09 \\
$R_{\rm A}$    (\Rsun) & 1.522    & 0.044    & 0.002     & 1.477 & 0.072       & \erc{1.440}{0.064}{0.057}    & 1.15 & 0.16 \\
$\log g_{\rm A}$ (cgs) & 4.189    & 0.021    & 0.001     & 4.169 & 0.033       & \mc{ }                       & \mc{ }      \\
$\rho_{\rm A}$ (\psun) & \mcc{$0.370 \pm 0.027$}         & 0.365 & 0.037       & \erc{0.394}{0.024}{0.032}    & \mc{ }      \\[2pt]
$M_{\rm b}$    (\Mjup) & 0.592    & 0.019    & 0.002     & 0.542 & 0.050       & \erc{0.551}{0.041}{0.038}    & \mc{ }      \\
$R_{\rm b}$    (\Rjup) & 1.408    & 0.046    & 0.002     & 1.428 & 0.077       & \erc{1.379}{0.067}{0.058}    & \mc{ }      \\
$g_{\rm b}$     (\mss) & \mcc{$7.39 \pm 0.46$}           & 6.08 & 0.62         & \mc{ }                       & \mc{ }      \\
$\rho_{\rm b}$ (\pjup) & 0.198    & 0.018    & 0.000     & 0.186 & 0.026       & \mc{ }                       & \mc{ }      \\[2pt]
\Teq\              (K) & \mcc{$1676 \pm   29$}           & 1652 & 28           & \mc{ }                       & \mc{ }      \\
\safronov\             & 0.0332   & 0.0013   & 0.0001    & \mc{ }              & \mc{ }                       & \mc{ }      \\
$a$               (AU) & 0.05165  & 0.00067  & 0.00008   & 0.0499 & 0.0018     & \erc{0.0499}{0.0019}{0.0017} & \mc{ }      \\
Age              (Gyr) & \ermcc{2.4}{0.6}{0.6}{0.2}{0.4} & \erc{3.9}{2.8}{1.3} & \mc{ }                       & \mc{ }      \\
\hline \end{tabular} \end{table*}

\begin{table*} \caption{\label{tab:16:model} Derived physical properties of the WASP-16 system and a comparison to 
previous measurements. Separate statistical and systematic errorbars are given for the results from the current work.}
\begin{tabular}{l r@{\,$\pm$\,}c@{\,$\pm$\,}l r@{\,$\pm$\,}l r@{\,$\pm$\,}l r@{\,$\pm$\,}l r@{\,$\pm$\,}l}
\hline
\ & \mcc{This work} & \mc{\citet{Lister+09apj}} & \mc{\citet{Doyle+13mn}} \\
\hline
$M_{\rm A}$    (\Msun) & 0.980    & 0.049    & 0.023     & \erc{1.022}{0.074}{0.129}    & 1.09 & 0.09 \\
$R_{\rm A}$    (\Rsun) & 1.087    & 0.041    & 0.008     & \erc{0.946}{0.057}{0.052}    & 1.34 & 0.20 \\
$\log g_{\rm A}$ (cgs) & 4.357    & 0.022    & 0.003     & \erc{4.495}{0.030}{0.054}    & \mc{ }      \\
$\rho_{\rm A}$ (\psun) & \mcc{$0.762 \pm 0.056$}         & \erc{1.21}{0.13}{0.18}       & \mc{ }      \\[2pt]
$M_{\rm b}$    (\Mjup) & 0.832    & 0.036    & 0.013     & \erc{0.855}{0.043}{0.076}    & \mc{ }      \\
$R_{\rm b}$    (\Rjup) & 1.218    & 0.039    & 0.009     & \erc{1.008}{0.083}{0.060}    & \mc{ }      \\
$g_{\rm b}$     (\mss) & \mcc{$13.92 \pm  0.71$}         & \erc{19.2}{1.9}{2.6}         & \mc{ }      \\
$\rho_{\rm b}$ (\pjup) & 0.431    & 0.033    & 0.003     & \erc{0.83}{0.13}{0.17}       & \mc{ }      \\[2pt]
\Teq\              (K) & \mcc{$1389 \pm   24$}           & \erc{1280}{35}{21}           & \mc{ }      \\
\safronov\             & 0.0579   & 0.0021   & 0.0004    & \mc{0.070 $\pm$ 0.010}       & \mc{ }      \\
$a$               (AU) & 0.04150  & 0.00070  & 0.00032   & \erc{0.0421}{0.0010}{0.0018} & \mc{ }      \\
Age              (Gyr) & \ermcc{8.6}{3.3}{2.7}{0.6}{0.9} & \erc{2.3}{5.8}{2.2}          & \mc{ }      \\
\hline \end{tabular} \end{table*}

The physical properties of the two systems can be determined from the photometric parameters measured from the light curves, the spectroscopic properties of the host star (velocity amplitude $K_{\rm A}$, effective temperature \Teff, and metallicity \FeH), and constraints from theoretical stellar evolutionary models. We used the approach presented by \citet{Me09mn},  which begins with an estimate of the velocity amplitude of the {\em planet}, $K_{\rm b}$. A set of physical properties can then be calculated from $K_{\rm A}$, $K_{\rm b}$, $r_{\rm A}$, $r_{\rm b}$, $i$ and orbital period using standard formulae. The expected radius and \Teff\ of a star of this mass and \FeH\ can then be obtained by interpolating within the predictions of theoretical stellar models. The value of $K_{\rm b}$ is then iteratively refined to maximise the match between the observed and predicted \Teff, and the measured $r_{\rm A}$ and predicted $\frac{R_{\rm A}}{a}$. This procedure is performed over a sequence of ages for the star, beginning at the zero-age main sequence and terminating once it becomes significantly evolved, in order to find the best overall fit and the age of the system.

We determined the physical properties of WASP-15 and WASP-16 using this approach, as implemented in the {\sc absdim} code \citep{Me09mn}, and the spectroscopic properties of the stars as summarised in Table\,\ref{tab:spec}. We have adopted the atmospheric parameters (\Teff\ and \FeH) from \citet{Doyle+13mn}, as this represents a thorough analysis of observational material of greater quality than for alternative measurements (see Sect.\,\ref{sec:intro}). The statistical errors were propagated through the analysis using a perturbation algorithm \citep{Me++05aa}, which has the advantage of yielding a complete error budget for every output parameter. 

Systematic errors are also incurred through the use of stellar theory to constrain the properties of the host stars; these were assessed by running separate solutions for each of five different sets of stellar model predictions \citep{Claret04aa,Demarque+04apjs,Pietrinferni+04apj,Vandenberg++06apjs,Dotter+08apjs} as implemented by \citet{Me10mn}. Finally, a model-independent set of results was generated using an empirical calibration of stellar properties found from well-studied eclipsing binary star systems. The empirical calibration follows the approach introduced by \citet{Enoch+10aa} but with the improved calibration coefficients derived by \citet{Me11mn}. The individual solutions can be found in Tables A10 and A11 in the Supplementary Information. We used the set of physical constants given by \citet[][their table\,3]{Me11mn}. 

Tables \ref{tab:15:model} and \ref{tab:16:model} contain our final physical properties for the WASP-15 and WASP-16 systems, plus published measurements for comparison. The mass, radius, surface gravity and density of the star are denoted by $M_{\rm A}$, $R_{\rm A}$, $\log g_{\rm A}$ and $\rho_{\rm A}$, and of the planet by $M_{\rm b}$, $R_{\rm b}$, $g_{\rm b}$ and $\rho_{\rm b}$. \Teq\ is the equilibrium temperature of the planet (neglecting albedo and heat redistribution) and $\Theta$ is the \citet{Safronov72} number. \reff{All quantities with a dependence on stellar theory have separate statistical and systematic errorbars quoted. The statistical errorbar for a quantity is the largest of the five errorbars found in the solutions using different theoretical model predictions. The systematic errorbar denotes the largest deviation between the final value of the quantity and the individual values from using the five different sets of models.}

\begin{figure} \includegraphics[width=0.48\textwidth,angle=0]{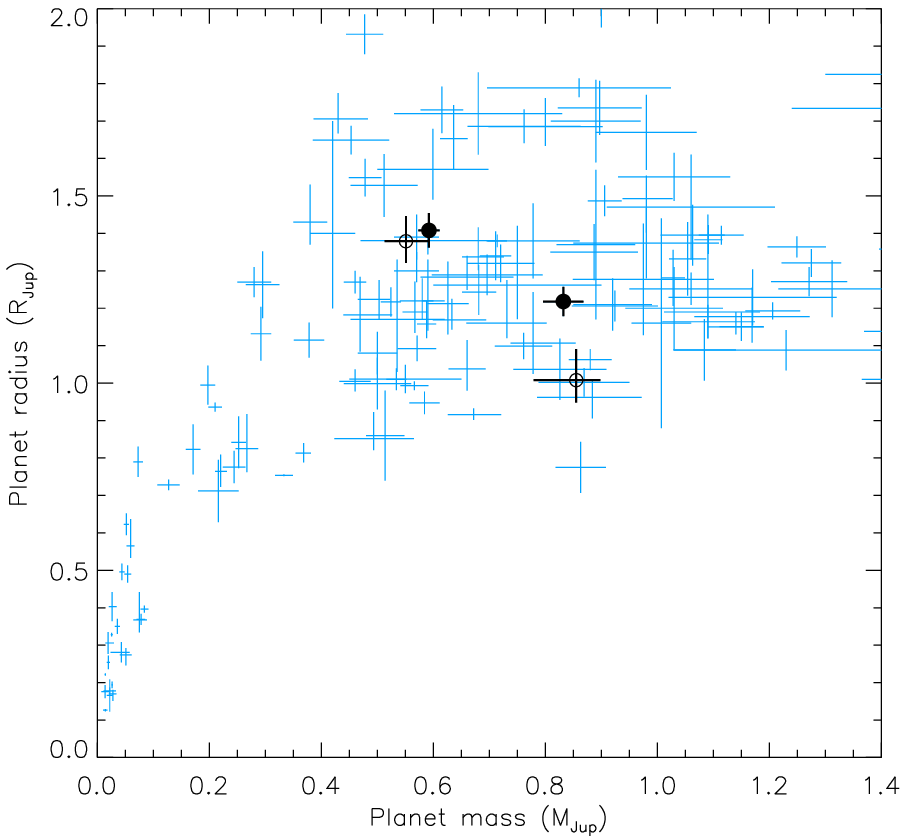}
\caption{\label{fig:m2r2} Plot of planet radii versus their masses. The 
overall population of planets is shown using blue crosses, using data taken 
from TEPCat on 2013/02/11. WASP-15\,b and WASP-16\,b are indicated using 
black lines with open circles for the values from their respective discovery 
papers and filled circles for the new results from the current work. The 
outlier with a mass of 0.86\Mjup\ but a radius of only 0.78\Rjup\ is the 
recently-discovered system WASP-59 \citep{Hebrard+13aa}.} \end{figure}

The higher \Teff\ adopted for WASP-15\,A in the current work caused us to find the star to be more massive and less evolved than previously thought. Our results are in good agreement with previous determinations but are significantly more precise due to the new photometry presented in this paper. Our results for WASP-16 go in the reverse direction: we find a less massive and slightly more evolved star (with a $\logg$ closer to the spectroscopic determination by \citealt{Doyle+13mn}). The planet WASP-16\,b is \reff{0.21\Rjup\ (2.5$\sigma$)} larger than previously thought, leading to a lower surface gravity and density \reff{by 2$\sigma$}. We find an old age of \er{8.6}{3.4}{2.9}\,Gyr for WASP-16, in agreement with the absence of emission in the calcium H and K lines (B.\ Smalley, private communication). The measurements for the planetary masses and radii are contrasted in Fig.\,\ref{fig:m2r2}.

\section{Summary}

WASP-15 and WASP-16 are two transiting extrasolar planets whose discovery was announced by the SuperWASP Consortium in 2009. Since then both have been the subject of follow-up spectroscopic analyses to measure their Rossiter-McLaughlin effects and host star temperatures, but neither have benefited from additional transit photometry to refine measurements of their orbital ephemerides and physical properties. We have rectified this situation by obtaining new light curves of two transits for WASP-15, of which one was covered in four optical passbands simultaneously, and of four transits for WASP-16. 

We modelled these photometric data using the {\sc jktebop} code with careful attention paid to limb darkening and error analysis, and augmented them with published spectroscopic parameters in order to find the physical properties of the components of both systems. Our approach followed that of the {\it Homogeneous Studies} project by the first author, and WASP-15 and WASP-16 have been added to the Transiting Extrasolar Planet Catalogue\footnote{The Transiting Extrasolar Planet Catalogue (TEPCat) is available at: {\tt http://www.astro.keele.ac.uk/jkt/tepcat/}}.

We confirm that WASP-15 is a highly inflated planet with a large atmospheric scale height which, when combined with the brightness of its host star ($V = 10.92$), makes it a good candidate for studying the atmospheres of extrasolar planets. Our simultaneous observations in four optical passbands are in principle good for probing this, so we attempted to do so using the methods of \citet[][]{Me+12mn2}. In practise, we found that our data are not extensive enough to allow inferences to be drawn. This is in line with previous experience \citep[][]{Mancini+13aa,Mancini+13mn,Nikolov+13aa} and could be rectified by obtaining new observations with GROND.

We find a significantly larger radius for WASP-16\,b, moving it from the edge of the mass--radius distribution to an area of parameter space more typical for transiting hot Jupiters. This underlines the point that multiple high-quality transit light curves are needed for the physical properties of a TEP to be reliably constrained. 
The detailed error budgets we have calculated show a typical situation: an improved understanding of both WASP-15 and WASP-16 would require additional transit light curves, radial velocity observations, and more precise \FeH\ determinations.


\section*{Acknowledgements}

This paper incorporates observations collected at the MPG/ESO 2.2\,m telescope located at ESO La Silla, Chile. Operations of this telescope are jointly performed by the Max Planck Gesellschaft and the European Southern Observatory. GROND has been built by the high-energy group of MPE in collaboration with the LSW Tautenburg and ESO, and is operated as a PI-instrument at the 2.\,2m telescope. We thank Timo Anguita and R\'egis Lachaume for technical assistance during the observations.
The operation of the Danish 1.54m telescope is financed by a grant to UGJ from the Danish Natural Science Research Council. 
The reduced light curves presented in this work will be made available at the CDS ({\tt http://vizier.u-strasbg.fr/}) and at {\tt http://www.astro.keele.ac.uk/$\sim$jkt/}.
J\,Southworth acknowledges financial support from STFC in the form of an Advanced Fellowship.
The research leading to these results has received funding from the European Community's Seventh Framework Programme (FP7/2007-2013/) under grant agreement Nos.\ 229517 and 268421. Funding for the Stellar Astrophysics Centre (SAC) is provided by The Danish National Research Foundation.
KAA, MD, MH, CL and CS are thankful to Qatar National Research Fund (QNRF), member of Qatar Foundation, for support by grant NPRP 09-476-1-078.
TCH acknowledges financial support from the Korea Research Council for Fundamental Science and Technology (KRCF) through the Young Research Scientist Fellowship Program and is supported by the KASI (Korea Astronomy and Space Science Institute) grant 2012-1-410-02/2013-9-400-00.
SG and XF acknowledge the support from NSFC under the grant No.\,10873031.
The research is supported by the ASTERISK project (ASTERoseismic Investigations with SONG and Kepler) funded by the European Research Council (grant agreement No.\,267864).
FF (ARC), OW (FNRS research fellow) and J\,Surdej acknowledge support from the Communaut\'e fran\c{c}aise de Belgique - Actions de recherche concert\'ees - Acad\'emie Wallonie-Europe.
The following internet-based resources were used in research for this paper: the ESO Digitized Sky Survey; the NASA Astrophysics Data System; the SIMBAD database operated at CDS, Strasbourg, France; and the ar$\chi$iv scientific paper preprint service operated by Cornell University.

\bibliographystyle{mn_new}

\end{document}